%% file: main.tex
\pdfoutput=1

\documentclass[prl,reprint,aps,superscriptaddress,nobibnotes,nofootinbib]{revtex4-2}
\usepackage[utf8]{inputenc}
\usepackage{mathptmx}
\usepackage{amsmath}
\usepackage{amssymb}
\usepackage[tight]{units}
\usepackage{color,graphicx}
\usepackage[colorlinks,citecolor=blue,urlcolor=blue,linkcolor=blue]{hyperref}
\usepackage[normalem]{ulem}
\usepackage{enumitem}
\usepackage{mathtools}
\usepackage{braket}
\usepackage{graphicx}
\usepackage{subfig}
\usepackage{makecell}

\usepackage[dvipsnames]{xcolor}

\begin{document}
%\begin{CJK*}{GB}{}
\title{Description of CRESST-II and CRESST-III pulse shape data}

\input{authors}
\begin{abstract}

A set of data from 68 cryogenic detectors operated in the CRESST dark matter search experiment between 2013 and 2019 was collected and labeled to train binary classifiers for data cleaning. Here, we describe the data set and how the trained models can be applied to new data. The data and models are available via Ref.~\cite{DMDC:CRESST}.
\end{abstract}
\maketitle
%\end{CJK*}

\section{Introduction}
\label{sec:intro}

Pulse shape processing is a standard task in the analysis of cryogenic detectors. The selection criteria for data cleaning used in the analysis chain are often hand-crafted for individual detectors. To reduce the amount of manual intervention in this process and to reduce human bias, neural networks have been trained to find universal selection criteria in Ref.~\cite{angloher_towards_2023}. For this project, a data set of pulse shapes and pulse shape features was collected with samples from all 68 detectors operated in the CRESST setup in three data-taking campaigns running from 2013 to 2019, dubbed run 33 to run 35. The total size of the data set is over 1 million records (voltage traces) and includes both background and calibration period data. We have released this data set to the community to enable the development and verification of pulse shape processing techniques. We have also released our pre-trained models that were characterized in Ref.~\cite{angloher_towards_2023}. 

The intended use of the released data is limited to pulse shape processing. We do not provide design information on individual detector modules or calibration sources, or other information that would be necessary to base a physics analysis (e.g. dark matter search) on these data sets. 

This document is organized as follows. We summarize the terminology used for the CRESST detectors in Sec.~\ref{sec:detectors}. We describe the data format of the released pulse shape data and pulse shape features in Sec.~\ref{sec:files}. Finally, we describe the data format of our pre-trained models as a Jupyter notebook for demonstrations in Sec.~\ref{sec:models}.

\section{CRESST detectors and analysis}
\label{sec:detectors}

The detector concept used by CRESST is based on transition-edge sensors (TES) coupled to absorber crystals and operated at about 15 mK. Details are given in Refs.~\cite{angloher_towards_2023, angloher2017descriptioncresstiidata, cresstcollaboration2020description}. The detector modules contain one or multiple, here up to three, absorber crystals that are equipped with TESs, each of which is recorded as a separate channel by the data acquisition system. For physics data taking these channels have individual purposes, e.g. detecting particle events or scintillation light, vetoing electromagnetic or vibrational backgrounds. Here we treat all channels individually and equally, ignoring coincidences between co-located sensors. The output of a sensor is a voltage trace, read out by a SQUID-based electronics which produces a digital signal with a sampling frequency of 25 kHz. 
%The SQUID output signal is split into two parts, where one is sent to a transient digitizer with a ring buffer, while the other part is sent to the trigger electronics, in which a comparator unit checks if a recorded sample exceeds the trigger threshold. In the case of a trigger, the DAQ waits for a pre-defined amount of time to read out the transient digitizer, including a set of pre-trigger samples, creating a record window around the triggered sample.
A particle recoil can be identified by its characteristic pulse shape. The readout electronics is triggered on the rise of the voltage trace above a pre-defined threshold, and a record window of either 8192 or 16384 values around the trigger site is stored for later processing. 
To monitor the noise, some records are randomly triggered without any channel exceeding the trigger threshold and are referred to as noise traces or baselines in this article. One run of the experiment typically includes one complete thermal cycle of the cryostat, a time period of several weeks or months. Changes in the configuration of the operated detectors and other hardware interventions/maintenance are performed between runs.

Parts of the data-taking periods of CRESST runs are dedicated to the energy calibration of the detectors. During these periods, the setup is exposed to strong $\gamma$-ray or neutron sources, typically dominating the total count rate in the data. 
Monoenergetic $\gamma$-ray or X-ray sources of known energy are used to calibrate the energy depositions in the absorber. The main purpose of the neutron calibration is determination of the positions and widths of the bands containing the different event classes in the energy-light yield plane, see Ref.~\cite{Angloher2024_likelihood} for details. For the data set described here, three measurement intervals of $\approx$50 h each were used as samples, with a $\gamma$/neutron/no source present in one of the intervals, respectively. We do not label which events originate from background or calibration data periods. 

% \FW{FILES TAKEN: run33: cal/2ndGamma/nr2_003, bck/bck_073, ncal2/ncal2_006; 
% run34: cal3/cal3_007, bg/bck_073, ncal1/ncal1_019; 
% run35: cal_nov18/cal_dec18_003, bck/bck_070, ncal/ncal_006}

The recorded pulse traces are characterized by pulse shape features such as pulse height, rise and decay times, baseline noise characteristics, and others. A detailed description of the pulse shape features used is listed in section ~\ref{sec:files}. The pulse height primarily carries the information on the energy deposited in the detector, with the pulse shape providing additional information in cases of saturation. The data cleaning process should remove all artifacts and events where the energy can not be properly reconstructed, while retaining all events that originate from individual energy depositions in the dynamic range of the detector and from particle interactions in the target.

The collected data set was labeled by defining acceptance regions in the pulse shape feature space for each detector individually. The meta-criteria for these cuts were described in Ref.~\cite{angloher_towards_2023}. They can be summarized as retaining all events that resemble the typical shape of pulses at the correct position within the record window, or empty noise traces, and rejecting all events that deviate from this. This manual cleaning procedure was performed using the Cait Python package \cite{wagner_cait_2022}. 

The data from 7 of the 68 detectors were designated as a test set and were excluded from the development of data cleaning methods, but were used only to test and compare their performance. The remaining 61 detectors are referred to as the training set. 
 
The performance of a detector is characterized by the noise level in units of equivalent energy deposition, called the baseline energy resolution. 
Here we estimate the baseline resolution by superimposing typical event shapes with known pulse height on a set of empty noise traces, filtering the resulting pulses with an optimum filter kernel, and reconstructing the spread in the pulse height estimation. Details of this method are described e.g.~in Ref.~\cite{wagner_towards_2023}. 
The filtering results in the best achievable pulse height estimate given a static pulse shape and stochastic noise frequency spectrum. 
The optimal filter resolution values reported in this data release are not given in units of equivalent energy deposition in the target. Instead, they are given in voltage units used to estimate the pulse height of the recorded pulses. Therefore, these values are not comparable between channels.

\section{Data files}
\label{sec:files}

The released data were converted from the original data recorded by the CRESST data acquisition system into formats accessible using standard Python libraries, such as NumPy \cite{harris2020array}, and text editors. Both raw voltage traces and derived pulse shape features, as well as additional information about the detectors and runs, have been released and are described below.

Two types of data files have been released: Numpy formatted memory-mapped numeric arrays (*.npy) and comma-separated values (CSV, *.csv). Memory-mapped Numpy arrays are written and loaded as described in the Numpy documentation (see also our demonstration notebook described in Sec.~\ref{sec:models}). The release includes four such files: 

\begin{figure}[t]
\centering
\includegraphics[width=\linewidth]{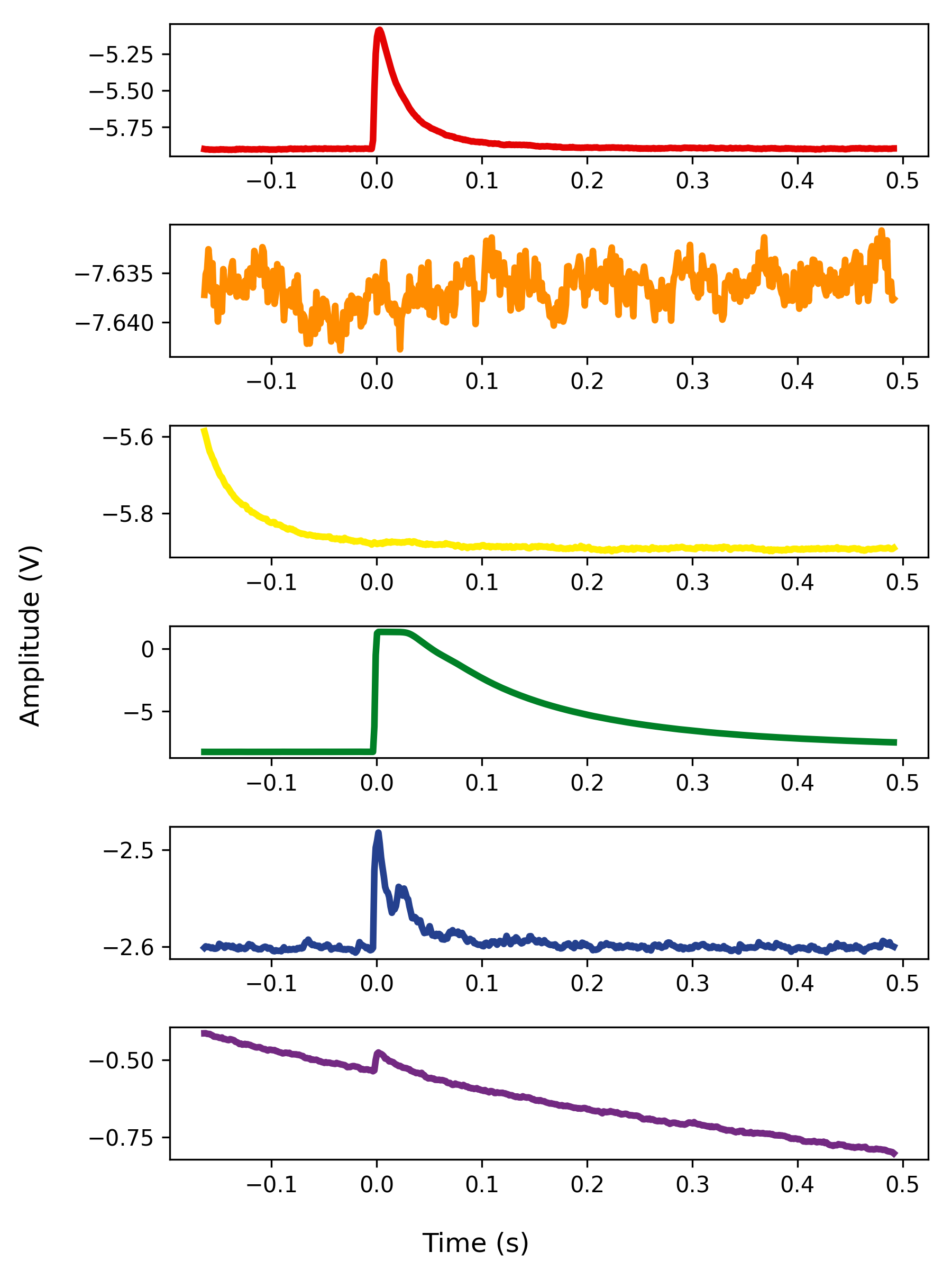}
\caption{Examplary records contained in X\_test.npy. The events are recorded at different offset voltages, which depend on the state of the readout electronics and have no physical meaning. The time is stated with respect to 1/4 of the record window (the default trigger position). Colors were chosen arbitrarily.}
\label{fig:pulses}
\end{figure}

\begin{itemize}

    \item \textbf{X\_train.npy} and \textbf{X\_test.npy} contain the raw voltage traces of all 979,446 records of the training dataset and 78,084 records of the test dataset, respectively. Each trace has a length of 512 32-bit float values. The records were downsampled from their original length of 8192 values in run 33, or 16384 values in runs 34 and 35, to contain 512 values. This was done by taking the average of 16 or 32 consecutive values. The X\_train.npy data was used in Ref.~\cite{angloher_towards_2023} as input to the neural network classifiers for training. Exemplary records from X\_test.npy are shown in Fig.~\ref{fig:pulses}. These records are the raw voltage trace from the output of the SQUID and hence have different baseline offsets.
    
    \item \textbf{y\_train.npy} and \textbf{y\_test.npy} contain integer values -1 or 1 corresponding to all 979,446 records of the training data and 78,084 records of the test data set. A positive value indicates that the record survived the quality cuts applied in our labeling scheme, while a negative value indicates that the record did not survive. These values were used as labels to train the classifiers in Ref.~\cite{angloher_towards_2023}. 

\end{itemize}

\noindent We have released four CSV files along with the raw data that contain derived features and additional characteristics of the records and other information:

\begin{itemize}

    \item \textbf{detectors.csv} contains information about the detectors used in the dataset. The rows correspond to the samples added to the dataset, in no particular order. The different columns and a description of their contents are given in Tab.~\ref{tab:detectors}. This file is intended to give an overview of the contents of the data set. It is not necessarily needed for training classification models or developing data processing methods, since all the event-by-event information and the information about the channel affiliation of the records is also written in the files described below. 

    \begin{table}
    \centering
    \begin{tabular}{ p{2cm}  p{6cm} } 
     \hline
     \textbf{Column} & \textbf{Description} \\ 
     \hline
     \hline
     run &  The number of the CRESST run (either 33, 34 or 35) from which the data sample was taken. \\ 
     \hline
     channel & The readout channel number within the run. \\ 
     \hline
     test &  If true, the data sample belongs to the test set. \\ 
     \hline
     nmbr\_events &  Number of records in the data sample. \\ 
     \hline
     nmbr\_positive &  Number of records in the data sample that survived the applied quality cuts. \\ 
     \hline
     noise &  If true, the data sample consists of induced random triggers, mostly containing noise traces. Pulses and artifacts can be visible on randomly triggered traces due to events coinciding with the record window by chance. \\ 
     \hline
    \end{tabular}
    \caption{Description of the columns in the detectors.csv file. Each row in the file corresponds to a data sample (a collection of records) that was added to the training or test set.}
    \label{tab:detectors}
    \end{table}

    \item \textbf{features\_train.csv} and \textbf{features\_test.csv} contain in their columns meta information, pulse shape features and the time stamp since the start of the data sample (the collection of consecutive records for the detector) for each record in the training and test datasets, respectively. The pulse shape features were calculated before downsampling the traces. The different columns and their descriptions are given in Tab.~\ref{tab:features}.  Each row in the file corresponds to the identically indexed sample in the training and test datasets, so the two files have 979,446 and 78,084 rows, respectively. 

    \begin{table}
    \begin{center}
    \begin{tabular}{ p{2cm} p{6cm} } 
     \hline
     \textbf{Column} & \textbf{Description} \\ 
     \hline
     \hline
     run &  The number of the CRESST run (either 33, 34 or 35) from which the record was taken. \\ 
     \hline
     channel &  The readout channel number within the run. \\ 
     \hline
     noise &  If true, the record was recorded due to an injected random trigger. \\ 
     \hline
     clean &  If true, the record survived the applied quality cuts. \\ 
     \hline
     pulse\_height &  The difference in V between the maximum value within the record and the offset (see below). \\ 
     \hline
     onset & The position within the record at which the pulse rises above 20\% of its maximal height, measured in milliseconds with respect to 1/4th of the record (the default trigger position). \\ 
     \hline
     slope &  The difference between the average of the first and last 500 values of the record before downsampling. \\ 
     \hline
     var &  The variance of all values within the record, after subtracting the offset (see below). \\ 
     \hline
     mean &  The mean value of all values within the record, after subtracting the offset (see below). \\ 
     \hline
     skewness &  The skewness of all values within the record, after subtracting the offset (see below). \\ 
     \hline
    array\_min & The minimum of all values within the record, after subtracting the offset (see below). \\ 
     \hline
     nmbr\_peaks & The number of peaks within the record, found with a Z-score peak finder algorithm similar to Ref.~\cite{brakel2014}.\\ 
     \hline
     max\_derivative &  The maximum of the derivative of the record, calculated as the numerical difference between subsequent values.  \\ 
     \hline
     min\_derivative &  The minimum of the derivative of the record, calculated as the numerical difference between subsequent values. \\ 
     \hline
     rise\_time &  The time in ms between the onset and the position within the record at which the pulse rises above 80\% of its maximal height. \\ 
     \hline
     decay\_time &  The time in ms between the position within the record at which the pulse falls below 90\% and 36\% of its maximal height. \\ 
     \hline
     offset & The mean value of the first 500 values within the record before downsampling. \\ 
     \hline
     hours & Time stamp of the trigger, measured in hours since the start of the data sample. \\ 
     \hline
    \end{tabular}
    \end{center}
    \caption{Description of the columns in the features\_train.csv and features\_test.csv files. All pulse shape parameters were calculated after calculating the 50 sample running average of the record.}
    \label{tab:features}
    \end{table}

    \item \textbf{resolutions\_test.csv} contains the optimal filter resolutions for the detectors used in the test set. The different columns and their descriptions are given in Tab.~\ref{tab:resolutions}. The optimal filter resolutions were used in Ref.~\cite{angloher_towards_2023} to describe the classifier accuracy in terms of signal-to-noise ratio.
    
    \begin{table}
    \begin{center}
    \begin{tabular}{p{2cm} p{6cm}} 
     \hline
     \textbf{Column} & \textbf{Description} \\ 
     \hline
     \hline
     run &  The number of the CRESST run (either 33, 34 or 35) in which the detector was operated. \\ 
     \hline
     channel &  The readout channel number within the run. \\ 
     \hline
     of\textunderscore resolution &  Optimum filter resolution in V. \\ 
     \hline
    \end{tabular}
    \end{center}
    \caption{Description of the columns in the resolutions\_test.csv file. Each row corresponds to one channel contained in the test data set.}
    \label{tab:resolutions}
    \end{table}
    
\end{itemize}

\section{Models}
\label{sec:models}

Along with the pulse shape data, we have released the four models trained and characterized in Ref.~\cite{angloher_towards_2023}. The file names and corresponding names of the models, identical to their naming in Ref.~\cite{angloher_towards_2023}, are written in Tab.~\ref{tab:models}. For user convenience, we have released a Jupyter notebook, \textbf{demo\_model.ipynb}, and two utility files, \textbf{utils.py} and \textbf{transformer.py}, demonstrating the correct usage of the models on the test dataset. The assumed folder structure for running the notebook is to have three same-level folders, ``data'', ``models'', and ``notebooks'', containing the data, models, and notebook, respectively. Note that the transformer model can only be loaded and used on machines with CUDA drivers. The other three models can be run on machines with CPU only. 

    \begin{table}
    \centering
    \begin{tabular}{p{5cm} p{3cm}} 
     \hline
     \textbf{Filename} & \textbf{Model} \\ 
     \hline
     \hline
     cnn\_aug\_fixed\_epoch\_14.model &  Convolutional Neural Network \\ 
     \hline
     lstm\_bidir\_aug\_fixed\_epoch\_14.model & Long-Short Term Memory Network \\ 
     \hline
     tscn\_aug\_fixed\_epoch\_15.model &  Time Series Convolutional Network \\ 
     \hline
     tst\_aug\_fixed\_epoch\_6.model &  Time Series classification optimized Transformer \\ 
     \hline
    \end{tabular}
    \caption{File names of released models and their corresponding names used in Ref.~\cite{angloher_towards_2023}.}
    \label{tab:models}
    \end{table}
    
The inputs to all models are the raw voltage traces of length 512. For correct usage, two additional preprocessing steps are required, which are also included in the Jupyter notebook as data transformations:
\begin{itemize}
    \item The values contained in each record window must be scaled so that the highest contained value is one and the lowest value is zero.
    \item The datasets must be converted to a PyTorch tensor~\cite{NEURIPS2019_9015}. 
\end{itemize}

\noindent The output of the models is a value between zero and one, and corresponds to the model's belief that the respective trace would survive the quality cuts in an analysis chain. To use the models for data cleaning, a threshold must be defined below which a data set is discarded. We recommend using a generic threshold of 0.5.

\section{Citation}

If you base your work on our data, we kindly ask you to cite this document as well as Ref.~\cite{angloher_towards_2023}. 

\section*{Acknowledgements}
We are grateful to LNGS-INFN for their generous support of CRESST. This work has been funded by the Deutsche Forschungsgemeinschaft (DFG, German Research Foundation) under Germany's Excellence Strategy – EXC 2094 – 390783311 and through the Sonderforschungsbereich (Collaborative Research Center) SFB1258 ‘Neutrinos and Dark Matter in Astro- and Particle Physics’, by the BMBF 05A20WO1 and 05A20VTA and by the Austrian Science Fund (FWF): I5420-N, W1252-N27 and FG1 and by the Austrian research promotion agency (FFG), project ML4CPD. The Bratislava group acknowledges a support provided by the Slovak Research and Development Agency (projects APVV-15-0576 and APVV-21-0377). 

\bibliographystyle{h-physrev}
\bibliography{main}
\end{document}

%% file: authors.tex
\newcommand{\mpi}{\affiliation{Max-Planck-Institut f\"ur Physik, D-80805 M\"unchen, Germany}}
\newcommand{\hephy}{\affiliation{Institut f\"ur Hochenergiephysik der \"Osterreichischen Akademie der Wissenschaften, A-1050 Wien, Austria}}
\newcommand{\ai}{\affiliation{Atominstitut, Technische Universit\"at Wien, A-1020 Wien, Austria}}
\newcommand{\lngs}{\affiliation{INFN, Laboratori Nazionali del Gran Sasso, I-67100 Assergi, Italy}}
\newcommand{\bratislava}{\affiliation{Faculty of Mathematics, Physics and Informatics, Comenius University, 84248 Bratislava, Slovakia}}
\newcommand{\tum}{\affiliation{Physik-Department, TUM School of Natural Sciences, Technische Universit\"at M\"unchen, 85747 Garching, Germany}}
\newcommand{\tue}{\affiliation{Eberhard-Karls-Universit\"at T\"ubingen, D-72076 T\"ubingen, Germany}}
\newcommand{\oxford}{\affiliation{Department of Physics, University of Oxford, Oxford OX1 3RH, United Kingdom}}
\newcommand{\coimbra}{\affiliation{Also at: LIBPhys-UC, Departamento de Fisica, Universidade de Coimbra, P3004 516 Coimbra, Portugal}}
\newcommand{\wmi}{\affiliation{Also at: Walther-Mei\ss ner-Institut f\"ur Tieftemperaturforschung, D-85748 Garching, Germany}}
\newcommand{\gssi}{\affiliation{Also at: GSSI-Gran Sasso Science Institute, I-67100 L'Aquila, Italy}}
\newcommand{\cassino}{\affiliation{Also at: Dipartimento di Ingegneria Civile e Meccanica, Universit\`a degli Studi di Cassino e del Lazio Meridionale, I-03043 Cassino, Italy}}

\newcommand{\kip}{\affiliation{Kirchhoff-Institut für Physik, Universität Heidelberg, 69117 Heidelberg, Germany}}

\newcommand{\saopaulo}{\affiliation{Also at: Instituto de Física da Universidade de São Paulo, São Paulo 05508-090, Brazil}}

\newcommand{\aeth}{\affiliation{Now at: Department of Physics, ETH Zurich, CH-8093 Zurich, Switzerland}}

\newcommand{\apsi}{\affiliation{Now at: ETH Zurich - PSI Quantum Computing Hub, Paul Scherrer Institute, CH-5232 Villigen, Switzerland}}

\mpi
\lngs
\tum
\hephy
\ai
\tue
\oxford
\bratislava
\kip

\coimbra
\gssi
\wmi
\cassino

\author{G.~Angloher} \mpi
\author{S.~Banik} \hephy \ai
\author{D.~Bartolot} \hephy
\author{G.~Benato} \lngs
\author{A.~Bento} \mpi \coimbra
\author{A.~Bertolini} \mpi
\author{R.~Breier} \bratislava
\author{C.~Bucci} \lngs
\author{J.~Burkhart} \hephy
\author{L.~Canonica} \mpi
\author{A.~D'Addabbo} \lngs
\author{S.~Di~Lorenzo} \lngs
\author{L.~Einfalt} \hephy \ai
\author{A.~Erb} \tum \wmi
\author{F.~v.~Feilitzsch} \tum
\author{N.~Ferreiro~Iachellini} \mpi
\author{S.~Fichtinger} \hephy
\author{D.~Fuchs} \mpi
\author{A.~Fuss} \hephy \ai
\author{A.~Garai} \mpi
\author{V.M.~Ghete} \hephy
\author{P.~Gorla} \lngs
\author{P.V.~Guillaumon} \mpi \saopaulo
\author{S.~Gupta} \hephy
\author{D.~Hauff} \mpi
\author{M.~Je\v{s}kovsk\'y} \bratislava
\author{J.~Jochum} \tue
\author{M.~Kaznacheeva} \tum
\author{A.~Kinast} \tum
\author{H.~Kluck} \hephy
\author{S.~Kuckuk} \tue
\author{H.~Kraus} \oxford
\author{M.~Lackner} \mpi
\author{A.~Langenk\"amper} \mpi \tum
\author{M.~Mancuso} \mpi
\author{L.~Marini} \lngs \gssi
\author{L.~Meyer} \tue
\author{V.~Mokina} \hephy
\author{P.~Murali}  \email[Corresponding author: ]{praveen.murali@kip.uni-heidelberg.de} \kip
\author{A.~Nilima} \mpi
\author{M.~Olmi} \lngs
\author{T.~Ortmann} \tum
\author{C.~Pagliarone} \lngs \cassino
\author{L.~Pattavina} \lngs \tum
\author{F.~Petricca} \mpi
\author{W.~Potzel} \tum
\author{P.~Povinec} \bratislava
\author{F.~Pr\"obst} \mpi
\author{F.~Pucci} \mpi
\author{F.~Reindl} \hephy \ai
\author{D.~Rizvanovic} \hephy
\author{J.~Rothe} \tum
\author{K.~Sch\"affner} \mpi
\author{J.~Schieck} \hephy \ai
\author{D.~Schmiedmayer} \hephy \ai
\author{S.~Sch\"onert} \tum
\author{C.~Schwertner} \hephy \ai
\author{M.~Stahlberg} \mpi
\author{L.~Stodolsky} \mpi
\author{C.~Strandhagen} \tue
\author{R.~Strauss} \tum
\author{I.~Usherov} \tue
\author{F.~Wagner} \email[Corresponding author: ]{felix.wagner@phys.ethz.ch} \hephy \aeth \apsi
\author{M.~Willers} \tum
\author{V.~Zema} \mpi

\collaboration{CRESST Collaboration}
\noaffiliation